\documentclass[twocolumn,runningheads]{svjour2}
\smartqed  
\usepackage{graphicx}
\journalname{Astrophysics and Space Science}
\begin{document}

\title{The Design of Diamond Compton Telescope
}


\author{Kinya Hibino        \and
        		Toshisuke Kashiwagi \and
         		Shoji Okuno \and
        		Kaori Yajima \and
        		Yukio Uchihori \and
        		Hisashi Kitamura \and
        		Takeshi Takashima \and
        		Mamoru Yokota \and
        		Kenji Yoshida
}


\institute{Kinya Hibino \and  Toshisuke Kashiwagi \and Shoji Okuno \and Mamoru Yokota  \at
               Faculty of Engineering, Kanagawa University, Yokohama, Japan\\
              Tel.: +81-45-481-5661\\
              Fax: +81-45-413-1437\
              \email{hibino@n.kanagawa-u.ac.jp}           
           \and
          Kaori Yajima \at
              Department of Physics, Toho University, Chiba, Japan
            \and
           Yukio Uchihori \and Hisashi Kitamura \at
           National Institute of Radiological Science, Chiba, Japan
          \and
          Takeshi Takashima \at
          Japan Aerospace Exploration Agency, Sagamihara, Kanagawa, Japan
          \and
          Kenji Yoshida \at
          Faculty of Systems Engineering, Shibaura Institute of Technology, Saitama, Japan
}

\date{Received: date / Accepted: date}

\maketitle

\begin{abstract}
   
We have developed radiation detectors using the new synthetic diamonds. 
The diamond detector has an advantage for observations of "low/medium" energy gamma rays as a Compton telescope. 
The primary advantage of the diamond detector can reduce the photoelectric effect in the low energy range, 
which is background noise for tracking of the Compton recoil electron.
A concept of the Diamond Compton Telescope (DCT) consists of position sensitive layers of diamond-striped detector 
and calorimeter layer of CdTe detector. 
The key part of the DCT is diamond-striped detectors with a higher positional resolution 
and a wider energy range from 10 keV to 10 MeV. 
However, the diamond-striped detector is under development.
We describe the performance of prototype diamond detector and the design of a possible DCT evaluated by 
Monte Carlo simulations.
   
\end{abstract}

\section{Introduction}
\label{intro}

The gamma-ray observation in the energy range from 10 keV to 10 MeV 
is crucial for the study of a rich variety of high-energy astrophysical process,
but it is hard pressed to observe gamma-rays in this energy range.
The Universe of keV - MeV gamma-ray astronomy has been opened up
by the COMPTEL telescope\cite{COMPTEL} on the Compton Gamma Ray Observatory satellite, 
which was launched in April 1991.
The COMPTEL telescope covered the energy from 750 keV to 30 MeV.
After that epoch, the INTEGRAL satellite, which was launched in October 2002, 
is taking up the observation of gamma-rays in the 20 keV to 10 MeV range.

The gamma-ray observation of the INTEGRAL satellite\cite{INTEGRAL-1}\cite{INTEGRAL-2} 
is very superior to it of
 the COMPTEL detector, but it is difficult
 to reach the standard like the current X-rays observation 
 under the present conditions.
Therefore, the next generation observatory has to go for more enhanced 
sensitivity by using new detector technology.

Diamond detectors have been developed mainly by Russian groups since 1970s\cite{DIA-1},
and it was shown that these detectors may have a performance which is 
equal to Si semiconductor detectors.
However, the performance of those detectors was unstable and it was difficult to expand the 
effective area.
Because conventional diamond detectors have been based on natural diamond, 
it was hard to acquire diamond material with a large area and 
uniform crystal.
Recently, a diamond material has been made artificially, with large area, uniform crystal and high degree of purity.

\section{Prototype diamond detector}
\label{sec:1}

\begin{figure}
\centering
  \includegraphics[width=0.2\textwidth]{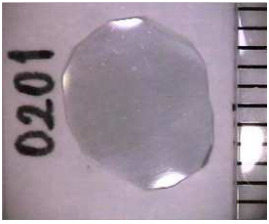}
\caption{A synthetic diamond which has 9 mm $\times$ 7 mm area and 183 $\mu$m.}
\label{fig:dia}       
\end{figure}

A prototype of synthetic diamond detector has been made of
high-purity type IIa diamond single crystals.  
Their crystals are synthesized using high-pressure and high-temperature method\cite{DIA-2}.
The typical sizes of the diamond crystals are about 3 $\times$ 3 mm$^2$ $\sim$ 10 $\times$ 10 mm$^2$, 
and the thicknesses are 100 $\sim$  300 $\mu$m and their uniformity is less than 0.5 \%, as shown in Fig.~\ref{fig:dia}.
The sensitive  area of the diamond detector is $\sim$ 1 mm diameter in our present detectors.

\begin{figure}
\centering
  \includegraphics[width=0.4\textwidth]{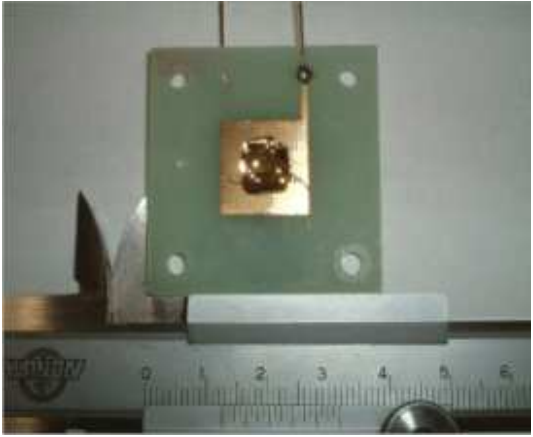}
\caption{Photograph of our synthetic diamond detector sample. A synthetic
diamond has been pasted on a glass epoxy plate by epoxy glue}
\label{fig:prototype}       
\end{figure}

\begin{figure}
\centering
  \includegraphics[width=0.4\textwidth]{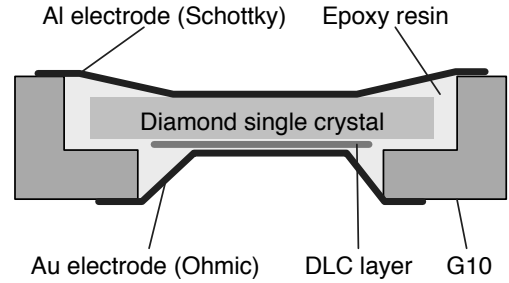}
\caption{Cross section of our synthetic diamond detector sample. Its surfaces
were put electrodes by vacuum deposition method}
\label{fig:crosssection}       
\end{figure}

Figure \ref{fig:prototype} shows one of  the prototype diamond detector and Fig.~\ref{fig:crosssection} 
shows a schematic cross section of our diamond detector.
The diamond plate was fixed on a glass epoxy board (G10) with epoxy adhesive.
Electrical contacts were prepared on each surface of the diamond plate by a vacuum evaporation method.
Evaporated aluminum contacted a diamond surface as a Schottky contact, 
and evaporated gold contacted an opposite diamond surface which was 
treated with a diamond like carbon (DLC) 
layer as an Ohmic contact.
The thickness of each electrode was roughly 20 $\sim$ 40 nm.

\section{Energy Resolution}
\label{sec:2}

The prototype diamond detector has been tested by using irradiation from some isotopes and heavy ion beam.

\begin{figure}
\centering
  \includegraphics[width=0.5\textwidth]{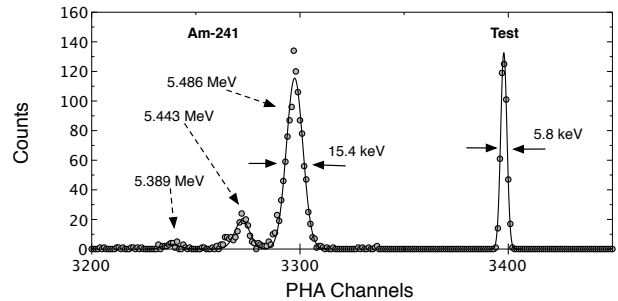}
\caption{The energy spectra of alpha-rays from $^{241}$Am obtained with our prototype diamond detector. }
\label{fig:energy_distribution}       
\end{figure}

The energy resolution was measured by alpha particles from $^{241}$Am source.
The energy distribution of our diamond detector is shown in Fig.~\ref{fig:energy_distribution}.
In this figure, peaks of 5.486 MeV,  5.443 MeV and 5.389 MeV for alpha particles from  $^{241}$Am source 
are clearly shown.
The energy resolution obtained is 15.4 keV (FWHM) for 5.486 MeV.
This result shows that the diamond detector has a comparable performance to a good silicon
detector which has a distribution spread of about 10 keV\cite{Knoll}.

\begin{figure}
\centering
  \includegraphics[width=0.5\textwidth]{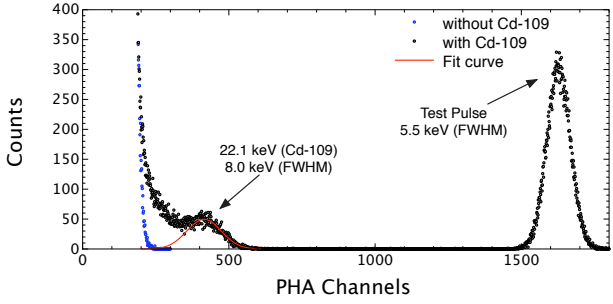}
\caption{The energy spectra of X-rays from $^{109}$Cd obtained with our prototype diamond detector. }
\label{fig:X_E}       
\end{figure}

Secondly, the diamond detector was irradiated by X-rays from $^{109}$Cd source. 
The energy distribution of our diamond detector is shown in Fig.~\ref{fig:X_E}.
A peak of 22.1 keV for X-rays from $^{109}$Cd is shown in this figure.
The energy resolution for X-rays obtained is 8.0 keV (FWHM) for 22.1 keV.
This result is not excellent in comparison with a typical silicon detector.
However,  it should be noted that no peak of more than 40 keV is measured at all.
This feature shows that the diamond detector is not influenced by photoelectric effect
because of low Z material.
This result suggests that the diamond detector has a wide range of Compton effect from low energy
in comparison with silicon detectors.

\section{Simulation}
\label{sec:3}

\begin{figure}
\centering
  \includegraphics[width=0.45\textwidth]{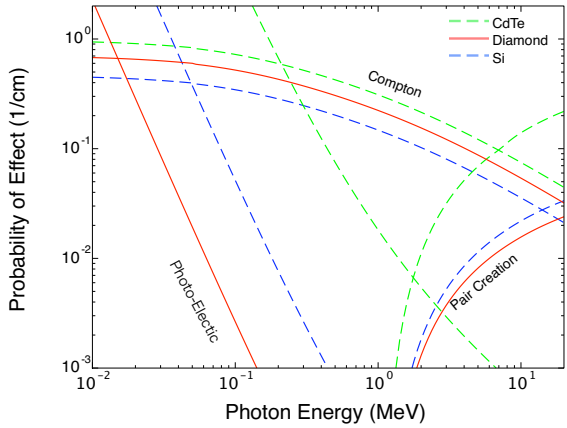}
\caption{Probability of various interactions in diamond, Si and CdTe as a function of photon energy.}
\label{fig:attenuation}       
\end{figure}

For an application of the diamond detector to a Compton recoil telescope, 
there is the advantage of a low Z material. The diamond material is not influenced 
by photoelectric effect as compared with Si and CdTe material. Therefore, the 
diamond detector is expected to work efficiently to detect gamma-rays with energy 
from 10 keV to 10 MeV through Compton scattering, as shown in Fig.~\ref{fig:attenuation}.

To determine the arrival direction and primary energy of a gamma-ray of this domain, 
it is necessary to measure correctly a track and an energy of electron by which Compton scattering is carried out 
and total energy of scattered gamma-ray. 
For that purpose, the basic detector design for DCT is composed of layers of diamond-striped detectors 
for position and energy measurement and a layer of CdTe detector for calorimeter.

The diamond-striped detector is still under development, and then, we examined the performance of 
DCT with our simulation. 
The used code of the Monte Carlo simulation is EPICS code\cite{cosmos}.

\begin{figure}
\centering
  \includegraphics[width=0.4\textwidth]{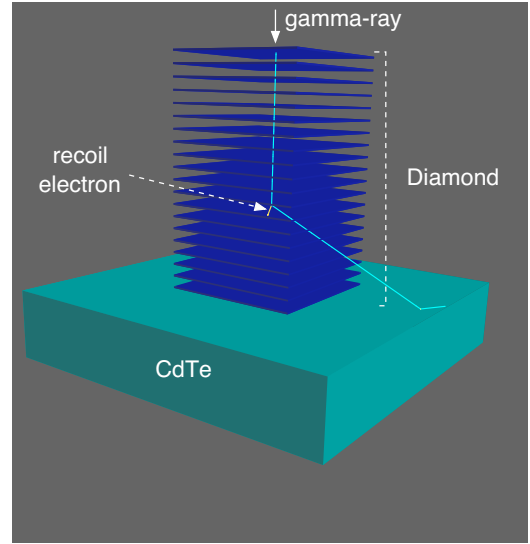}
\caption{The example of a simulated event for the gamma-ray of 500 keV 
which caused Compton scattering in the diamond layer once.}
\label{fig:dcrt}       
\end{figure}

The model of DCT used for the simulation consists of 20 layers of diamond of 4 cm $\times$  4 cm $\times$  0.5 mm 
 for tracking measurement of recoil electrons and the block of CdTe of 10 cm $\times$ 10 cm $\times$  4 cm as a calorimeter layer.
Figure \ref{fig:dcrt} shows an outline of the detector, and shows the example of a 
simulated event for a gamma-ray of 500 keV which caused Compton scattering in the 
diamond layer once. 
Detection efficiency of this prototype 
model was calculated. To determine the arrival direction from single or 
multi Compton scattering, it must be scattered around in diamond layers and
must stop in a CdTe layer. 

\begin{figure}
\centering
  \includegraphics[width=0.5\textwidth]{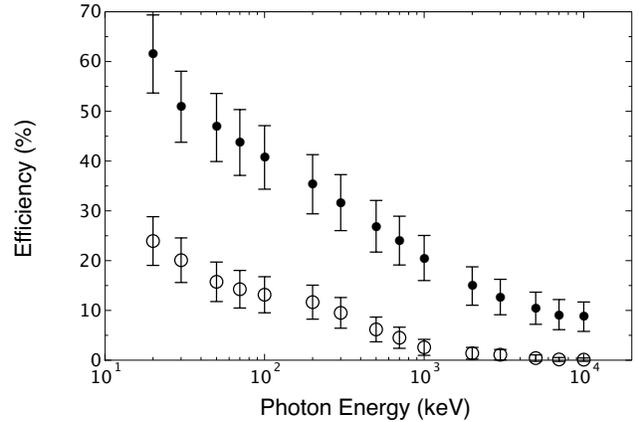}
\caption{The peak detection efficiency of 20 layer diamond detectors by using Monte Carlo simulation. 
Filled circles indicate the event ratio which carried out Compton scattering about two or more times, 
and open circles indicate event ratio which it was scattered about and stopped in the CdTe block.}
\label{fig:efficiency}       
\end{figure}

Figure \ref{fig:efficiency} shows the peak detection efficiency in the 
case of the 20 layer diamond detector by using by Monte Carlo simulation, which 
is about 24 \% at 20 keV and about 13 \% at 200 keV. 
From this result, it became clear that 
a diamond detector  use can be for gamma-rays observation above 10 keV. 
However, it turns out that about 60 \% or more of gamma-ray of all events will 
be scattered around upward in this model. 
This result shows that the detection efficiency could be improved further, 
when preparing the CdTe layer with a shape surrounding diamond layers.

\section{Conclusions}
\label{sec:4}

Development of radiation detector made natural diamond has been done since 1970s.
Because, the strong binding energy of diamond results in hardness and stability that 
can withstand harshly thermal, chemical and radiation environments. 
Furthermore, the wide band gap energy of diamond ($\sim$ 5.5 eV) leads to low leakage current and 
low sensitivity to visible light.
Diamond characteristics  are excellent as a material of the detector.

We have developed a semiconductor detector by using synthetic diamond crystal.
The prototype detector characteristics have been investigated with a radiation source,
and has shown many advantages which are comparable with a good Si detector.
And then, we have shown a possible application of DCT in the low energy band (10 keV $\sim$ 10 MeV).
To demonstrate the advantages of the DCT, we plan to develop a prototype of diamond-striped detector
and compare the results with more detailed simulations to optimize the design for DCT.

\begin{acknowledgements}
This research was supported in part by Grants-in-Aid 
for Scientific Research (A), 13354002, 2001-2004 and 
Grant-in-Aid for Young Scientists (B), 14740172, 2002-2004
 from the Ministry of Education, Culture, Sports, Science and 
Technology of Japan(MEXT).
\end{acknowledgements}



\end{document}